# Data management for platform-mediated public services

*Challenges and best practices*

# Author(s)

*Agnieszka Rychwalska, University of Warsaw,*

*a.rychwalska@uw.edu.pl*

*Geoffrey Goodell, University College London, g.goodell@ucl.ac.uk*

*Magdalena Roszczynska-Kurasinska, Uniwersity of Warsaw,*

*magda.roszczynska@gmail.com*

**Abstract**

Data harvesting and profiling have become a *de facto* business model for many businesses in the digital economy. The surveillance of individual persons through their use of private sector platforms has a well-understood effect on personal autonomy and democratic institutions. In this article, we explore the consequences of implementing data-rich services in the public sector and specifically the dangers inherent to undermining the universality of the reach of public services, the implicit endorsement of the platform operators by government, and the inability of members of the public to avoid using the platforms in practice. We propose a set of good practices in the form of design principles that infrastructure services can adopt to mitigate the risks, and we specify a set of design primitives that can be used to support the development of infrastructure that follows the principles. We argue that providers of public infrastructure should adopt a practice of critical assessment of the consequences of their technology choices.

**Introduction**

Services mediated by ICT platforms, including car sharing, hotel booking, social media, and more, have shaped the landscape of the digital markets and catalyzed immense economic opportunities. Unfortunately, the growth of platform-mediated services comes with a cost. One of the most serious challenges is the pervasive surveillance of individual interactions with the services that leads to accumulation and centralization of "Big Data": linked transactions and attributes related to individual persons. The users of platforms not only surrender the value of their digital traces (i.e. their privacy is compromised) but also subject themselves to the power and control that data brokers exert for prediction and manipulation (Zuboff 2015).

As the platform revolution takes hold in the area of public services, it is important to first safeguard such services from the pitfalls that have already been identified in the application of platform services to other areas. Just as importantly, public services have a responsibility to ensure that their technology is accessible, suitable, and appropriate for everyone. In pursuit of innovation, many policymakers embrace technologies without proper consideration of all the costs and risks. Misuse of Big Data collected via the administration of public services might be harmful not only because the data might be particularly sensitive and detailed but also because the users might not have a legitimate choice about whether to share their data.

In this paper we explore beyond the deleterious consequences of surveillance for the privacy of citizens: we identify the crucial, potentially devastating ways in which data from surveillance can be used to bypass democratic processes. Rather than choosing a single definition of a democratic regime, we follow Diamond and Morolino (2004) and focus on certain dimensions of political processes, such as trade-offs involving representation and accountability, within a society and analyze how data-rich services can alter the position of a polity on these dimensions.

Additionally, we propose a set of design principles for data systems in public services, which can serve as a guideline or benchmark in the assessment and deployment of platform-mediated services.

Finally, we propose a set of generic and generative design primitives that can be used to fulfil the proposed constraints and exemplify best practices in the deployment of platforms that deliver services in the public interest. We suggest that policymakers could adopt these design primitives and best practices as standards by which the appropriateness of candidate technology platforms can be assessed in the evaluation of their suitability for delivering public services.

**Platform revolution and democracy**

The last decade has seen a revolution in the digital market. The terms "sharing economy" (Hamari, Sjöklint, and Ukkonen 2016; Sundararajan 2016), "platform economy" (Kenney and Zysman 2016), and "platform revolution" (Choudary, Van Alstyne, and Parker 2016) all describe similar models that rely on so-called *collaborative consumption* (Belk 2014; Botsman and Rogers 2010). However, the true key to the success of platform-based services is collection, aggregation, and analysis of data that allow their proprietors to sell advertisements or fitting products to customers. This platform surveillance constitutes a new economic form that defines the dominant practices of online service providers (Wood and Monahan 2019).

The data gathered through such "dataveillance" (Clarke 1988) and the resulting profiling of users can be abused not only for the financial profit of platform owners at the expense of their customers (Zuboff 2015) but also by other actors to target persuasive social and political messages. Much public debate has been devoted to the possibility of the harm such practices deal to democratic processes (Badawy, Ferrara, and Lerman 2018; Allcott and Gentzkow 2017; Aral and Eckles 2019; Helbing et al. 2019). Democracy in this understanding (Diamond 2019) relies on the informed, fact-reliant choices of citizens that are not predominantly driven by emotional response or herding behavior, both of which can be induced by tailored messages.

Because the term "democracy" is applied to increasingly differing political systems (Mechkova, Lührmann, and Lindberg 2017), it may be tempting to overlook the impact of such abuses of data accumulation and treat them as a new form of political or social campaigning. It rests in the hands of national and international law makers to decide to what extent this behavior should be limited and to take appropriate action. We argue here that if similar platform and app-based, data-rich services would be implemented in the public sector without appropriate, *a priori* regulation, for example to furnish smart city lighting, voting, or personalized assistance in any public service, the problem could be exacerbated. This is even more problematic when the provision of the services is outsourced to specific actors in the private sector, creating a state-corporate nexus (Hayes 2012) with tightly-coupled responsibilities and accountability. The private sector can abuse data by bypassing laws or taking advantage of loopholes and inadequate law ("Cookiebot Report: Hidden Tracking of Citizens on EU Government and Health Sector Websites" 2019). However, if governmental agencies abuse collected data, then such abuse could become a legitimized practice, leaving little to no avenue for appeal.

The scope of democratic liberties and procedures varies greatly among nation states, and it is not our goal to analyze what difference data gathering and misuse could make in each case. Instead we would like to highlight how the systematic aggregation of data, centralization of data, and "nudging" within personalized services can progressively shift the balance between values that are recognized as inherent to democratic political systems, such as freedom, equality, and control (Bühlmann et al. 2012; Bochsler and Kriesi 2013). Values are characteristics towards which individuals and groups strive, some of which are by definition contradictory, creating moral dilemmas that need to be resolved (Schwartz 2012).

The defining principle of democracy is that power is assigned to a broad set of decision-makers (e.g. Knutsen 2010; Campbell 2008). While many different implementations of this principle can be traced through history and among current democratic states, varying in the degree and form of devolution, the common point is that democratic polities recognize the ideological and pragmatic value of pluralism, such as better representation of local conditions, better adoption of policies or greater adaptability (Manor 1999; Ribot 2007). Diversity is recognized as conducive to better solutions and adaptation in various sciences, including political science (Ober 2008), social science (Cioffi-Revilla 2005), complexity (Simon 1962), systems science (Page 2008), and many others.

The advantages to increasing pluralism among voices that decide on the course of a social system, such as a nation state or an institution, are counterbalanced by the effectiveness of the decision-making process. Achieving a working consensus is easier when there are fewer voices to reconcile (Arrow 1951). In the extreme case decision making is the easiest when there is a single voice. This is due not only to the fact that combining inputs is harder when there are more of them, but also to the fact that including more voices means accepting that more of them will not be well-informed or good-willed.

We can thus think of representative democracy as a way to optimize between the advantages of pluralism and the effectiveness of singular decision makers (Dahl 1994). Effective representation requires that those chosen to represent are accountable to those that have chosen them (Diamond and Morlino 2004). To ensure such accountability, the power to "control" is given both to the governing as well as the governed (Bühlmann et al. 2012). Such mutual accountability might sacrifice some short-term effectiveness for long-term stability; for example, in communities governing pooled resources, accountability is one of the principles that protect them from depletion (Ostrom 1990).

Another value that is typical in modern democracy but stems from the Athenian model of democracy, is *privacy*, specifically in the sense that no citizen has a right to interfere with the doings of others (Goldschmidt 1954). Privacy may be considered crucial to maintain independence of voices. When voices become uniform, the advantages of pluralism disappear (Page 2008). Similarly to the case of pluralism, increasing privacy carries costs. In particular, some measure of scrutiny might be necessary for collective security. This balance relies on mutual accountability among decision-makers, which in turn requires strong laws or norms to prevent the most powerful from avoiding their obligations to the rest.

In this paper we aim to analyze how the implementation of particular, implicit practices related to data, many of which are already entrenched in the private-sector services, can affect the balance between democratic values. We also present design principles that can be used to consciously choose and establish the point of balance. We posit that awareness of the trade-offs of data based services should be a requirement in policy decisions.

**Platform revolution in the public sector**

*Massive data gathering*

If digital platforms were implemented in the public sector to provide public services via platforms and apps, then the Big Data gathered from user behavior and from various device sensors could provide the basis for revolutionizing the public sector. Services would be better fitted, policy-making could become more data-driven and evidence-based, and the costs would be reduced for both the administration of the public sector as well as for citizens, embodying Giddens' vision of welfare surveillance wherein social supervision translates into better public services (Weller 2012). This would comply with a widely shared sentiment that while Big Data advantages have boosted the private sector, similar advantages have yet to be realized in the public sector (González-Bailón 2013; Kim, Trimi, and Chung 2014). However, extensive data gathering and analysis in the public sector can lead to the same dangers of manipulating public opinions and behavior as have been identified in the private sector (Helbing 2016), even more so as many public services rely specifically on sensitive data.

So far, reservations about Big Data and related "dataveillance" (Clarke 1988) in the public sector were usually focused on security issues (Mergel, Rethemeyer, and Isett 2016) and privacy (Desouza and Jacob 2017). It is worth noting that commitment to securing the data once gathered is tantamount to a belief that both current and future data custodians will not be willing to abuse their access (Aste and Goodell 2018). Moreover, data protection and regulation of data *use* do not, by themselves, imply preserving individual privacy: privacy necessitates minimizing and regulating data *collection* as well (Nissenbaum 2017). We note in particular that once data are collected, it is impossible to prove to a data subject that the data have not been copied, preserved, or otherwise misused.

The consequences of "dataveillance" transcend breach of personal privacy and national security and involve the fundamental mechanisms built into democratic societies, such as the decentralization and accountability that underpin the balance between the values of efficiency and pluralism, and between the values of freedom and control.

The first reason for inducing such imbalances is the fact that the act of observing citizens through the lens of sensors and metrics from apps and devices can change the behavior of the observed (Landsberger 1958), both through incentives as well as through chilling effects (Schauer 1978). An example comes from China, whose government started testing the usefulness of a social credit system for "increasing social trust" by incentivizing behaviors the government deems trustworthy ("Planning Outline for the Construction of a Social Credit System (2014-2020)" 2014). At first look, the motivation for this system might seem to be to build a Foucaultian panopticon, wherein citizens, aware of being observed, self-govern and change their behavior (Elmer 2012). However, any data gathering that is used to incentivize certain behaviors also leads to classification: surveillance as sorting (Gandy Jr 2012).

Even for benevolent governments, an existential goal is to stay in power, and statistical "boxing" can be employed as a bureaucratic tool of control (Bowker and Star 2000) to reduce accountability. In democratic societies the principle of equality (Diamond and Morlino 2004) requires that public services should be available and accessible to all citizens in a non-discriminatory manner. Any classification can simplify exclusion, upsetting the balance between the equality and control dimensions of effective democracy (Bühlmann et al. 2012). For example, data gathering itself, as a result of its dependence upon technology adoption and deployment, can be biased to exclude some groups, for instance elderly citizens due to low technology use and fluency, or citizens of poorer regions due to low public investment in technology in these places (Desouza and Jacob 2017; Mergel, Rethemeyer, and Isett 2016; *Reuters* 2014). In effect, the services themselves will evolve to suit selected segments of the population, whilst for other segments, the access to services, as well as the control over how these services are implemented and assessed, will be limited.

Moreover, such statistical surveillance disproportionately affects populations that are already at a disadvantage. Just as data surveillance in commercial platforms causes cumulative disadvantages for the poorer members of society (Gandy Jr 2012), public services targeted at marginalized populations often purportedly require increased surveillance (Monahan 2017). What is especially alarming is that the nature of statistical surveillance as remote, distant (Gandy Jr 2012), and mechanistic, contributes to dehumanizing narratives about the marginalized populations (Monahan 2017). This happens even in situations in which the purported motivation of introducing such citizen observation is to improve their welfare.

When the assumption of benevolence on the part of platform and service providers is violated, segmentation of citizens can be used to specifically inconvenience certain groups, favor certain others or simply ignore minorities once they are identified. Such actions might mistakenly attributed to the specific design of the technology: Excuses such as "technological inevitability" are commonly used to cover deliberate negligence (Zuboff 2019). Similarly, the term "platform", when used to describe a service, connotes neutrality and obscures the agency of the technology supplier (Plantin et al. 2018) in the pursuit of its own interests (Langlois and Elmer 2013), irrespective of whether it is a public agency or a contractor from the private sector that provides the service.

Classification enabled by amassing data and incentivizing behaviors within services can thus shift the balance between the effectiveness of services and the accountability of the public agencies providing the service. Yet another danger might lie in the aspects of this system that diverge from a panopticon scenario: by being unobtrusive and obscure in its goals and motivations, it can even more surely and imperceptibly alter citizen behavior in the process of behavioral nudging.

*Data aggregation and behavioral nudging*

While IT surveillance leads to massive data gathering, this by itself, and even combined with easily unpacked incentives for certain behaviors, is a starting point for much more complex applications when it is used as a source for prediction and covert (rather than based on clear incentives) modification of behaviors. Just as data are used in commercial platforms, data gathered on individual users ("citizen-consumers") can feed algorithms that can both predict the behavior of individuals as well as control such behavior through appropriately chosen and placed cues. Behavioral "nudging" (Sunstein and Thaler 2003; Sunstein 2014) can be more successful than direct "pushing" (i.e. incentives) for the reason that because its mechanics are opaque and difficult to understand, it cannot be easily opposed (Helbing et al. 2019).

The aim of nudging is to change citizen behavior for the greater good, i.e. into pro-social behavior as defined by the policy-makers, through small cues embedded in the environment of the public space (Shafir 2013; Olejniczak and Śliwowski 2015). An example is the placement of speed bumps on roads in residential areas or of roundabouts in dangerous spots on roads with higher speed limits to force drivers to slow down.

Historically, environmental affordances have been used to push questionable policies. For example, certain black and white neighborhoods in the U.S. were separated with highways to limit the possibility of contact by increasing the required effort (Lessig 2000). Architectural constraints are effective for such subversive goals, as they are often perceived by citizens as an inevitability rather than a planned choice and are thus overlooked.

If data from many services and many citizen actions are aggregated and linked via Big Data analytics into detailed user profiles, the cues can be tailored to the characteristics of particular individuals on the basis of their vulnerabilities, thus reducing the chance for informed choice (Zuboff 2019). Pervasive surveillance of citizens can thus be more than a mere breach of privacy, it can also undermine the agency of citizens through imperceptible control of their behavior. In effect, both the balance between privacy and control as well as the balance between effectiveness and accountability may shift.

Although unlimited data amassing by itself can upset the balance, even greater dangers come from linking the various records in the databases, either directly, such as to each other or to the identity of the data subject, or indirectly, for example by the time, location, or type of an activity. For individuals, the power to unilaterally link together different attributes or transactions associated with an individual person threatens the right to privacy. If multiple requests for the service (e.g. wherein persons board public transportation vehicles) can be linked to each other (e.g. the same passenger card was used), then the person can be profiled, and his or her linked activities can provide additional information (e.g. the person's occupation or employer) even when other identifying information is not known (O'Hara, Whitley, and Whittall 2011). Further examples of inferences might include determining an individual's home address by identifying travel starting points, identifying habits and social class from destinations in shopping and dining areas, and so on. Such aggregation and linkage of data points (while possibly making the service more effective) infringes on the privacy of citizens – privacy requires not only protecting precise identification in the form of a name or an ID, but ensuring that the complex, intertwined characteristics, behaviors, and associations of a given person remain within that person's control.

On the system level, the detailed profiles of citizens, irrespective of how they are linked (perhaps by a personal ID, or by some other feature, such as a device MAC address or a debit card number), create profound vulnerabilities that can be abused to compromise democratic processes. From targeted campaigns, to systematic neglect of specific citizen groups, to suppression of the rights of political opponents, to manipulation of needs and opinions, the linkage of data and profiling of citizens may become a powerful tool of control (Helbing et al. 2019).

Taken together, data gathering and behavioral nudging based on data can destabilize certain self-regulatory mechanisms inherent to democratic regimes. Massive data gathering on a vast fraction of citizens can change the behavior of the observed especially if it allows segmentation of the population and the development of different incentives for different groups. Such a process can undermine the pluralism of opinions and behaviors in the population. Algorithms operating on the data can add to this the ability to personalize the cues and triggers for certain behaviors with unprecedented accuracy while remaining unseen, hiding this process from public oversight, undermining accountability of the representatives, and possibly limiting sustainability (e.g. of public resources).

*Data centralization*

One of the principles of democracy is the assignment of decision-making power to the citizens. In representative democracy this principle is implemented by decentralization of governance, i.e. a process of decision making that underscores the variability in local circumstances, situation, needs and values (Manor 1999).

The main risk in a free flow of data between government authorities and public services within and across departments and institutions lies in the fact that data shared in this way contributes to building a centralized data cache. While the opportunities from having such a cache are often described (Desouza and Jacob 2017), we argue that this process upsets the balances between the control of the government and control of the constituents. The surveillance of citizens through IT solutions involves extracting value from behavior (Plantin and Punathambekar 2019) and centralization of data caches moves this value away from local representatives. In this manner, the centralized database moves power from local, decentralized departments to centralized decision making.

Moreover, centralization of data from different sources can enable linking of records on individual persons, even if they do not poses the same unique identifiers. Specifically, the overlap of specific traits and characteristics is enough to aggregate data (O'Hara, Whitley, and Whittall 2011).

In contrast, local databases, organized in an application-specific way, with unlinked user data, may provide local authorities with social capital to improve the local conditions by developing fitting solutions with public funds at their disposal. Public services such as transportation, public space, cultural artifacts, and so on can be enhanced by improving their functioning, accessibility, or display based on locally-gathered data.

In this way, the knowledge that local institutions can bring to the table when negotiating with their peers or with the central government is unique and constitutes social capital, a resource that grows and evolves in a path dependent way, inseparable from local history and values, and which cannot be reduced to a certain number of bits of data. Transferring data to a central cache detracts from the usefulness of this capital, as the role of such knowledge in policy making on national level will be diminished.

A centralized cache may create an impression that local governance and local institutions are no longer needed to represent local needs, as their needs are believed to be represented in the data. With data centralized, the stakeholder set is reduced to whoever manages the database (presumably central government) and individual citizens tasked with securing their needs, leading to a significant concentration of power. Individuals are forced to interact with actors who are not interested in their particular needs (rather, of the society as a whole) and whom they have little power to influence (Dahl 1994). This shifts the balance between the power to control of the governed and of the governing (Bühlmann et al. 2012), limiting the former's ability to hold public institutions accountable.

The problem of data centralization is becoming more and more complex with the sophistication of digital services, which most often requires that public agencies need to outsource the implementation, maintenance and analytics to private contractors. The interplay of public and private sector (i.e. the state-corporate nexus, Hayes 2012) presents a daunting challenge to ensure that citizen rights are preserved. For example, companies providing US police with IT solutions (e.g. cameras and recording devices and tools to analyze the data) try to optimize their financial interests by improving their products and services for their immediate customer, the police departments (Gates 2019), potentially leading to simplification of procedures or sharing of sensitive data that are convenient for police officers but bypass citizen rights.

This is even more problematic with the current drive to offer software as a service rather than a product. In this model, the front end of applications serves only as an interface for users whilst most of the functions – including data storage and analytics – are performed in the cloud, on the servers and databases of the service provider. This further complicates the processes of accountability of governmental and state agencies that aim to deliver the service but subcontract its provision to private sector. Each use of such service will create many copies of data, as the mere act of transfer between clients and servers will create copies. This is not only a security issue, but also regulatory problem: data required for the services might become centralized without proper recognition by the public bodies officially providing the service.

In effect, centralized databases holding citizen data reduce both the representation of diversity as well as the decentralization of governance, while possibly increasing speed and efficiency of the decision making process.

*Example: The SARS-CoV-2 Pandemic*

The complexity of the balances and trade-offs in data handling in the public sector is perfectly exemplified by the different ways in which central governments implemented technologies, applications, and platforms to contain the SARS-CoV-2 virus. For example, in Russia, face recognition is used to enforce quarantine orders, in Poland selfie-posting is used for the same purpose, and in South Korea apps are used to track social contacts of citizens (Seerat 2020).

In the case of pandemic, the value of privacy is juxtaposed against the value of public health. In principle, massive data gathering could help trace contacts between the infected and the susceptible and thus not only save lives and health of those primary contacts but also prevent further infections and reduce the spreading of the virus. Centralization of such data, i.e. combining them into a single database on the level of the nation state or federation of states, might enable the modelling and prediction of the epidemic spread. If data were combined across many sources (such as data from other apps that have information about medications, sports activities, general lifestyle, or potential medical treatment), the aggregate dataset could even be used to inform studies on the virus and disease itself, which could serendipitously help find a cure or a vaccine.

However, such serendipity comes at a high price, as pervasive long-term tracking of the mobility of individuals is an invasion of privacy. This can be further exacerbated if also other data are collected, such as social relations, health related behaviors or medical history. Not only privacy is compromised: such sensitive data can be used to construct detailed distributions of health behaviors and health condition of citizens, for example to classify them into risk categories, allowing insurance firms to capture consumer surplus. Furthermore, the more data and the more detailed they are, the easier it would be to statistically draw conclusions about illnesses, habits, previous medical treatments, family status, and employment status beyond the specific fields in the database. Further, when amassed and aggregated this data could provide a vast trove of information for targeting political campaigning, addressing in a personalized way the ailments of each individual. If the curators of such database were benevolent, this might simply boost their political position. However, if this assumption were violated, then such analytics could be used to exclude or discriminate against the vulnerable, limiting costs by preferentially providing health care to those in better condition, or to discriminate those who request politically sensitive medical treatments that are ideologically opposed by the curator of the data or otherwise stigmatized (e.g. abortion, in vitro procedures, repeated addiction therapy, etc.). Moreover, if the database were accessed without permission, its contents could be exploited and misused by various organizations, such as insurance companies, credit score assessors, employers, etc.

The construction of a centralized, linked database to address emergencies related to the pandemic could bring not only a multitude of advantages but a multitude of risks as well. If the choice is set in this way, it might be a difficult decision. It should not be the case that the price for health security is paid in democratic or human rights. Certain limitations on citizens' privacy and freedom are needed in times of an emergency, but as the EU Commissioner for Human Rights, Dunja Mijatovic stated, these can be employed "only under exceptional and precise conditions while offering adequate legal safeguards and independent supervision" (Mijatovic 2020). The conditions she lists, including but not limited to anonymity, encryption, non-discriminatory design, and transparency, are analogous to the checks and balances that are characteristic of democratic systems and required for accountability (Diamond and Morlino 2004).

Fulfilling these requirements can be greatly simplified by adhering to some general design principles. In the case of disease tracking these would include decentralization of data storage (e.g. on the users device instead of cloud), opening of the source code for transparency and user control, limitation of data gathering (e.g. only a time sorted list of anonymous, single use identifiers for physical space contacts, rather than the precise locations of users or personally identifying information about their contacts), the prevention by design of data linkage between different apps and different uses of the same app (e.g. restricting access to device-specific identifiers, employing single-use identifiers that cannot be cross-identified as belonging to the same user or to other specific users within the same location), and so on. In the next section we describe these principles in more detail and generalize them for other possible services.

**Design principles for data management in the public sector**

The transfer of public services from offline into the digital by implementing them in mobile applications and web-based platforms can improve the participation of citizens and increase convenience. However, this process can upset the balances between various dimensions of democratic procedures and as such should be carefully monitored. It is not enough to simply state that public services should not contribute to the surveillance of citizens, as any digital service will inevitably collect data. Rather, researchers, engineers and policy practitioners should focus on establishing good practices to guide the process of digitization of public services.

*Principle 1. Minimizing data collection*

Our first and foundational principle is that public services delivered via applications and platforms and in principle all services that digitize data on their users, should strive to collect only the minimal amount of data required to supply the service. Note that this approach is fundamentally different from "data protection", or data use regulation which assume that a broad set of data will be collected by service providers and the effectiveness of which relies upon their policy adhesion, compliance procedures, and enforcement.

The principle of minimizing data collection has additional caveats. First, the decision on what data is truly needed should not be determined by the service providers themselves, especially not on the basis of "technological inevitability". We argue that neither the technology itself not its low cost justify data gathering. There is a push to innovate in the public sector, but innovation or implementation of innovative technologies should not be a goal in itself. Rather, technology should be a means to achieve goals set by or negotiated with the constituents.

*Principle 2. Data decentralization*

Data decentralization ensures that privacy rights integral to democratic polities can be recognized and balanced properly. The fact that digitalized data on citizens can be shared in a costless manner does not justify breaking such balances. Nor does "technological inevitability" associated with automatic synchronization of data, automatic transfers, or backups that involve services or entities beyond the original supplier of service justify upsetting the balance. We argue that exactly because the transfers are so easy and do not incur costs they should be policed even more vigilantly.

While in certain circumstances some degree of data centralization might be necessary, this process should be carefully regulated and subject to oversight by appropriate public institutions. For example, metadata from contact tracing in the case of an epidemic, as described earlier, might require some centralization. Similarly, intelligence operations might benefit from centralized databases. However, all such cases require oversight from dedicated public bodies: much abuse results from insufficient checks and balances on such procedures of data accumulation (Snowden 2019). While it might be true that certain inferences are possibleonly when data are gathered at scale (e.g. White et al. 2013), this proposition implies that the power to make such inferences is limited to those that have access to a massive, centralized database and capabilites to analyze the data, thus shifting the balance of power.

We argue that public institutions should be held to the same or higher standards as private ones when it comes to passing data to third parties, including not only private-sector contractors but also public institutions and governmental agencies beyond the specific one that the citizen is sharing data with within a particular service.

Whenever the creation of a technology behind a service or the provision of the whole service is outsourced to private companies or non-governmental organizations, these entities should be held to the same principles. Specifically, they should be required to operate within the regulatory perimeter and to be subject to public oversight just as public institutions are. This includes transparency in the information that is collected and how it is used, justification for the necessity of data collection and requirement of explicit consent when collecting and sharing data. Moreover, citizens should not be forced into non-consensual relationships with third party operators. At minimum, when provision of the service is impossible without outsourcing to private agents, the citizens' freedom of choice of private-sector providers should be protected. A selection of providers should be available so that the users are able to assess their trustworthiness and to develop trust relations at will.

We posit that public sector institutions should be responsible for the whole process of data gathering, analysis and use in the service: they should be able to track, control and take responsibility for every part of the process, even when parts of it are subcontracted to private actors. This can be done e.g. through dictating unitary terms of service that would oblige the direct service provider, as well as any third parties involved to follow good practices in data management. Moreover, citizens should be able to easily appeal any mishandling of data, with clear points of contact and accountability. Vigilance and expertise is required to regulate and monitor digital public services so that the private sector does not abuse its position as the supplier of services.

*Principle 3. Keeping data linkages a citizen prerogative*

We argue that the practice of allowing users to control the linkages among their data records should be part of the design in all public services provided digitally. Citizens should be aware not only of the risks of unlimited data linking but also of their rights to remain private, and their responsibility in keeping the data secure. Ultimately, more responsibility will lie with the individual. The challenges involved in the process of providing citizens with the knowledge, skills and competence required to protect their data should not be a reason to forego proper education and promotion of data management on the level of individuals. Historically, many technological inventions (including cars and personal computers) have been described as too difficult for the general public (or certain stereotyped groups) to operate, which has not prevented their wide adoption. In turn, this would require public service providers to organize proper education and informational campaigns.

Ensuring that ordinary persons can keep their data secure shall require an effort not only on the part of citizens but also on the part of= central and local governments and non-governmental organizations. First, any data-rich public services should be deployed following best practices, such as the design principles presented here. The user interfaces should be designed with care, so that non-critical data are not collected by default. It is also crucial that citizens understand that data collection of this sort can endanger the democratic processes, by manipulation of opinions, reduction of pluralism, and evasion of accountability. Protection of privacy can be construed as a public good problem: while some individuals might not care for their own privacy and might be willing to share their data freely, a critical mass of citizens protecting their data is necessary to mitigate the risks of data accumulation described earlier. Thus protection of one's data can be viewed as one of the civic responsibilities of citizens of democratic states.

**Techniques**

We identify five specific techniques that can be used as design primitives for the establishment (and, in many cases, refactoring) of technology infrastructure for public services. The main objectives of these techniques are to avoid the collection and aggregation of personal data and to keep the linkage between data records pertaining to individual persons strictly in the hands of the individual persons themselves.

*Technique 1. Metadata-resistant, attribute-based credentials.*

The first technique we propose can serve the first principle – minimizing data collection – as well as the last one – allowing individuals to manage the linkages between their data. The idea is to minimize gathering unnecessary data attributes of service users and to rely on the specific attributes that are required to give access to a service. Offline credentials and many digitized signed credentials contain metadata such as biometrics (e.g. photographs), references to other credentials (name, address, etc.), or simply open-ended fields that can be populated with arbitrary information. Sometimes these metadata are automatically transferred when a third party credential is used to authenticate an individual. This can happen for example, when a user logs in with a social media login to be able to comment on a public service website, or when social media profiles are used for public services.

Metadata included in these ways are self-defeating, particularly for electronic credentials that are automatically associated with transaction records when they are presented, if our goal is to minimize data collection and prevent linkages between transactions carried out by the same individual. Rather than providing issuers with a way to potentially de-anonymize their clients, systems should instead require issuers to generate a set of signing keys, one for each attribute for which the issuer can make an attestation about an individual, and use the appropriate key to generate a bare signature without any metadata. The particular attribute implied by the signature is determined from the choice of key that is used to generate the signature. For example, if citizens paying their taxes in a particular region are eligible for reduced fees for public transportation services in that region, the only attribute that they should be required to present to receive the discount should be "Y is a tax payer in region X". Thus, a key specific to attributes of tax paying locations should be used to sign such a message which can then be shown to public transportation officials to authorize the citizen to receive reduced fares. No additional information should be required or automatically linked. It may seem that the easiest way would be for the citizen to present a paper or a digital tax form that he or she submitted. However, this way the citizen inadvertently reveals many more personal attributes, such as name, residential address, income, marital status, and so forth. Instead, she should be able to receive a signed credential from the tax office that does not include any other attributes than the office where the tax form was submitted.

A significant challenge to personal privacy arises from the perception of strong non-transferability as a desirable property of credential systems. Digital identity systems, therefore, often rely upon the assumption that individuals will establish a single 'root' identity or master key that links together all of the various credentials that they would seek to attest. In fact, many and perhaps most public infrastructure services do not actually require strong non-transferability. Consider for example access to library resources, public transportation, and mobile data services, each of which can be provided by governments or private-sector institutions alike. Goodell and Aste have demonstrated that it is possible to design a system of bearer credentials that can facilitate use of such services without exposing individuals to the risk of profiling by exploiting provable relationships among the credentials (Goodell and Aste 2019).

*Technique 2. Blind signatures.*

A related technique that would support attribute based authorization are blind signatures. This solution can also help minimize data collection and automatic linkages. If the issuer of a credential signs a message about a client, and if the client then delivers this message to a relying party or to anyone else, then the issuer will be able to identify and track the signature when it is furnished to a relying party. This creates a problem, particularly if the issuer is asked to maintain identifying records linking its clients to its signatures and to disclose information on demand that links the signatures to those records. The solution is for the system to support blind signatures. A blind signature scheme requires a 'blinding' function that (a) when applied to a message, obfuscates the contents of that message for anyone who does not know the inverse of the function, and (b) commutes with the cryptographic signing function that will be used by the signer of some message (Chaum 1983). Then, a client can first apply the 'blinding' function to some message, then send the 'blinded' message to an issuer to be signed, and finally apply the inverse of the 'blinding' function to reveal a signature on the original message. The effect is akin to (a) writing a message on a slip of paper, (b) inserting that slip of paper along with a slip of carbon paper into a sealed security envelope, (c) asking the issuer to sign the outside of the envelope and then return the envelope without breaking the seal, and finally (d) opening the envelope to reveal the original message with the signature of the issuer.

The value of this approach is that, although everyone can see that the issuer signed the envelope, the issuer is not able to recognize the unblended signature when it is presented. In this manner, the issuer would not be able to maintain records that could subsequently de-anonymize the client. By obviating such an attack, the use of blind signatures aligns the interests of the issuer with the interests of its clients. Furthermore, because the issuers would beless valuable as targets of powerful adversaries, they would have reason to support and encourage this approach.

This technique has applications in anonymous voting procedures, but it can just as easily be applicable to any digital service where only a selected number of attributes is required to authenticate or authorize a user. Eligibility to vote can be verified based on nationality, residence or similar attributes by appropriate bodies without the need for them to know how the vote was cast. Similarly, the eligibility to pay reduced fare in public transport can be verified by tax offices without the public transportation knowing how much a citizen earns. Blinded signatures thus allow the service receiving such credentials to verify the user's eligibility without knowing anything else about the individual.

*Technique 3. Single-use credentials.*

Credentials such as standard-issue ID numbers or credit cards – or even online login data, especially if a single account is used for many services -  that can be used many times are considered convenient for users because they do not require sophisticated devices and online interactions. Unfortunately, this convenience comes at the expense of privacy as observers associate multiple transactions with the same identifier and use this information to construct a profile. Potential observers include not only service providers who are able to identify repeated visits from the same user, but also third parties such as platform operators and governmental agencies beyond the provider of the service. We suggest that the linkage problem is the inevitable result of credential reuse, and the solution is to avoid credential reuse altogether by using single-use credentials: an individual presents a particular credential to a service provider exactly once and then never uses the same credential again. Single-use credentials share important features with fungible tokens; features that allow tokens to be distinguishable from one other based on their transaction history might impact their value in use (Berg 2019).

 Physical, single use tickets are perhaps the simplest form of single-use credentials. Individuals purchasing tickets for public transportation do not generally need to reveal any information about themselves, and when they validate the ticket or present it to vehicle operators, no information is attached except that they are authorized to board the vehicle. The same principles should be obeyed if public transportation tickets, and any other credentials, are implemented in digital form.

*Technique 4. User-generated identifiers and keys.*

For citizens to manage the linkages between their data, they must be able to generate their own identifiers. When third parties such as the manufacturers of devices or identity cards choose identifiers or keys on behalf of their users, then users have no way to be sure that there is no link between the identifiers or keys and the users that can be exploited (Abelson et al. 1997). Such exploitation could be carried out either by the third parties themselves, by their business partners, or by others with the power either to monitor their activities, to coerce them to disclose information about specific linkages, or to introduce weaknesses into the mechanism that they use to generate identifiers or keys. Users should not accept identifiers or keys that might be linkable. The solution is for users to generate identifiers and keys on their own devices, using well-audited, open-source technology. Although this step requires a degree of capability on behalf of the user's devices as well as a significant public responsibility to conduct audits and maintenance activities, we suggest that these steps are essential.

*Technique 5. Distributed infrastructure.*

Finally, we argue that public services should be run on distributed rather than centralized infrastructure. Infrastructure operated by a global platform operator is problematic at a system level for several reasons, including lack of diversity, lack of incentive for evolution, the opportunity for rent seeking, and so on. Also, the central platform operator has a chance to observe the metadata about all of the transactions, even if the transactions themselves are encrypted. For example, although the global operator of an encrypted chat platform might not be able to read the messages that its users send to each other, the global operator would still be able to see who talks to whom, when, how often, and for how long. Such metadata can still be used to precisely target individuals (Healy 2013). Most importantly, the central operator could change the rules of the system without the need to publicly ask the system operators to change their systems to follow the new rules. This friction, inefficient as it may be, is an essential check on the power of central authority, even when it has the legitimate power of oversight.

**Practical considerations**

Identification of design principles and candidate techniques that can fulfil the chosen constraints on data management is the first step to implement privacy-preserving and systemically robust digital public services. However, successful implementation requires that these principles are clear to all stakeholders and broadly adopted. To ensure their adoption and to well-target any changes we need to identify the group of stakeholders that is both in position to implement appropriate design principles as well as has interest in doing so.

This group includes executive officials, often specialists, tasked with procuring, managing and overseeing the implementation of specific policies and public services, including their digital versions.

These experts, who generally work on behalf of the executive branches of the government, are not driven by short election cycles, are bound by more direct relations of trust and accountability to the constituents, and can benefit from appropriate implementation of data management. This group could benefit greatly from targeted knowledge bases and best practice reviews that promote a critical approach to data management – including outsourcing to commercial organizations – in the public sector. For this reason, executive officials are especially valuable to policy dialogue, wherein scientific concepts and analysis of the full consequences of digitizing the public sector can meet the practical constraints of policy making. Such dialogue could be further enhanced by creation of national and international organisations to establish guidelines, standardized assessments, and technological audits of digitalized public services.

**Acknowledgments**

Geoffrey Goodell would like to thank Professor Tomaso Aste, the Centre for Blockchain Technologies at University College London, and the Centre for Technology and Global Affairs of the University of Oxford. The work presented in this paper was supported by the Engineering and Physical Sciences Research Council (EPSRC) through the BARAC project (EP/P031730/1), the European Commission through the FinTech project (H2020-ICT-2018-2 825215), and the Polish National Science Centre through grant No. 2017/27/B/HS6/00626.